\documentclass[referee]{raa_rb}           
\usepackage{graphicx,times}
\usepackage{natbib}
\usepackage{amssymb,amsmath}
\bibpunct{(}{)}{;}{a}{}{,}

\usepackage[a4paper=true,dvipdfm=true,pagebackref=true]{hyperref}
\hypersetup{pdftitle = The title of my PDF, pdfauthor = My name, pdfsubject= The subject, pdfkeywords = keyword1 keyword2 keyword3} 
\hypersetup{colorlinks = true, linkcolor = green, anchorcolor = red, citecolor = blue, filecolor = red, pagecolor = red, urlcolor = red}

\begin{document}

\title{High-Resolution Spectroscopy of Winds Associated with
T Tauri Stars
}

 \volnopage{ {\bf 2015} Vol.\ {\bf X} No. {\bf XX}, 000--000}
   \setcounter{page}{1}

   \author{Naoto Iguchi\inst{1}, Yoichi Itoh\inst{2}}

   \institute{Graduate School of Science, Kobe University, 1-1 Rokkodai-cho, 
Nada-ku, Kobe, Hyogo 657-8501, Japan  \\
        \and
             Nishi-Harima Astronomical Observatory, Center for Astronomy, 
             University of Hyogo, 
             407-2, Nishigaichi, Sayo, Hyogo 679-5313, Japan; 
             {\it yitoh@nhao.jp} \\
\vs \no
   {\small Received 2015 July 12; accepted 2015 xx xx}
}

\abstract{
We carried out optical high-resolution spectroscopy of T Tauri stars
using the Subaru Telescope. 
Using archive data of the Keck Telescope and the Very Large
Telescope, we detected forbidden lines of [S II] at 4069 \AA,
in addition to those of [O I] at 5577 \AA~ and 6300 \AA, for
13 T Tauri stars.
We consider that low-velocity components of these forbidden lines emanate from
the wind associated with T Tauri stars.
Using two flux ratios of the three lines,
we simultaneously determined the hydrogen density and temperature of the winds.
The winds of T Tauri stars
have the hydrogen density of $2.5\times10^{6}$ cm$^{-3} - 2.5\times10^{9}$
cm$^{-3}$ and the temperature of 10800 K -- 18000 K.
The mass loss rates by the wind are estimated to lie in the range
from $2.0\times10^{-10}$ M$_{\odot}$ yr$^{-1}$ to $1.4\times10^{-9}$ 
M$_{\odot}$ yr$^{-1}$.
The mass loss rates are found to 
increase with increasing mass accretion rates.
The ratio of the mass loss rate to the mass accretion rate is 0.001 --
0.1 for the classical T Tauri stars and 0.1 -- 1 for the transitional
disk objects.
}

   \authorrunning{N. Iguchi and Y. Itoh}            
   \titlerunning{Wind of YSOs}  
   \maketitle

%
\section{Introduction}           
\label{sect:intro}

An outflow is an energetic phenomenon associated
with a T Tauri star (TTS).
It removes angular momentum from the system, accelerating the
stellar evolution.
The outflows observed in the optical and infrared wavelengths
are classified into the following two groups (\citealt{Hartigan95}).
The first is a jet, which is a high-velocity component of the outflow.
It is well collimated and often extends hundreds of astronomical unit (AU).
The jet emits forbidden lines in the optical and near-infrared wavelengths, 
such as the [O I] line at 6300 \AA, the [S II] line at 6716 \AA, 
the [S II] line at 6731 \AA,
and the [N II] line at 6586 \AA.
The emission line of the jet is blue-shifted about 200 km s$^{-1}$ 
relative to that of the
central star and has a velocity width of $\sim$ 150 km s$^{-1}$.
The second component of the outflow is called wind.
It is considered that the wind emanates from an inner portion of
a circumstellar disk,
though its launching point has not yet been spatially resolved.
It is not well-collimated and the motion of the gas is slow.
Thus, the wind exhibits a low-velocity component of the lines.
It also emits forbidden lines, such as the [O I] line at 6300 \AA~
and the [O I] line at 5577 \AA.
In the millimeter wavelengths, molecular outflows are also detected 
toward many TTSs.

The hydrogen density and temperature of the gas are fundamental
parameters of the outflow.
\cite{Bacciotti99} calculated the electron density, 
temperature, and
ionization fraction of a jet, using the fluxes of the [S II] 6716,
[S II] 6731, [N II] 6583, and [O I] 6300 lines.
Because both [S II] 6716 and [S II] 6731 lines are emitted from 
the same excitation level, the flux ratio of these lines depends only on 
the electron density of the jet.
After deriving the electron density, they calculated the temperature
and ionization fraction of the jet 
using the flux ratios of the [N II] 6583, [O I] 6300,
and [S II] 6731 lines.
The estimated hydrogen densities and temperatures range between $10^{3}$
and a few $10^{4}$ cm$^{-3}$ and between 9000 and 12000K, respectively.

\cite{Hartigan95} estimated the mass loss rates of the jets
from the luminosities of the [O I] 6300 emission lines.
They assumed an electron density of the jets of $7 \times 10^{4}$ cm$^{-3}$ 
for all TTSs, and calculated the masses of the jets.
By multiplying the velocity by the mass and dividing by the length of the
jet, they concluded that the mass loss rates
are between 10$^{-8}$ M$_{\odot}$ yr$^{-1}$ and 10$^{-10}$ M$_{\odot}$
yr$^{-1}$.
Based on spectrophotometry in blue region, Gullbring et al. (1998)
estimated the mass accretion rates of 29 classical TTSs typically between
$10^{-9}$ and $10^{-7}$ Msun yr$^{-1}$. With these values,
\cite{Cabrit02} claimed that
the ratios of the mass loss rates
to the mass accretion rates range from 0.01 to 1 for TTSs.

\cite{Kwan95} constructed a model of TTS wind, in which
the mass loss rate is expressed as a function of 
the electron density and temperature of the wind.
The forbidden lines with a critical density higher than
$10^{6}$ cm$^{-3}$ are considered to be emitted in the wind region.
The critical density of the [OI] line at 5577 \AA~ is $1.3\times10^{8}$
cm$^{-3}$ and that of the [OI] line at 6300 \AA~ is $1.4\times10^{6}$
cm$^{-3}$, thus these lines are considered as of wind origin.

However, the electron density and temperature of the wind have
not been simultaneously determined from observations.
The ratios of the flux of [O I] 5577 to that of [O I] 6300 for the
low-velocity components are between 0.1 and 1
(\citealt{Hartigan95}).
The ratio of 0.1 corresponds to a temperature of 5000 K if the electron density
is 10$^{8}$ cm$^{-3}$, or a temperature of 15000 K if the electron density
is 10$^{6}$ cm$^{-3}$.
With only two emission lines, one cannot determine the density and
temperature of the wind simultaneously.

In addition to the [O I] lines at 5577 \AA~ and 6300 \AA,
we investigate the [S II] line at 4069 \AA.
Its critical density is $6.9\times10^{6}$ cm$^{-3}$
. We assumed that the low velocity components of the 
[S II] line at 4069 \AA~ and the [O I] lines at 5577 \AA~ and 6300 \AA~
are signatures of a disk wind.
By using three emission lines, the density and temperature can be
determined simultaneously.

\section{Observations and Archive Data}

We carried out optical high-resolution spectroscopy of 7 TTSs
using the High Dispersion Spectrograph (HDS) mounted on the Subaru Telescope.
The targets are single TTSs with spectral types of late K or early M.
The data were obtained on 2012 September 8, with the StdTd mode and
the 0\farcs6 width slit. 
This setting of the instrument achieved the wavelength coverage of 4020 -- 6780
\AA~ with a spectral resolution of $\sim$ 60,000.
The integration time for each object was between 600 s and 1600 s.

We also used archive data for nine TTSs obtained using the High Resolution
Echelle Spectrometer (HIRES) mounted on the Keck Telescope.
The spectral types of the targets are between K5 and M3.5.
The data were acquired by G. J. Herczeg on 2008 January 23, 
G. W. Marcy on 2008 May 23 and 2008 December 3,
S. E. Dahm on 2010 February 4,
and T. E. Armandroff on 2010 December 2.
The spectral ranges covered the [S II] line at 4069\AA, the [O I] line
at 5577\AA, and the [O I] line at 6300\AA.
The spectral resolution was $\sim$ 70,000.
The integration time for each object was between 300 s and 1800 s.
In addition, we used archive data of field dwarfs with spectral 
types identical to those of the TTSs.

Archive data for two TTSs obtained with
Ultraviolet and Visual Echelle Spectrograph (UVES) mounted on the VLT were
also used.
The data were acquired by N. E. Piskunov on 2000 April 16 and 
H. C. Stempels on 2002 April 19.
The spectral resolution was $\sim$ 80,000.
The integration times were 1800 s and 1900 s.
Dwarf spectra acquired by C. Melo on 2009 April 2 were also used.
The observed TTSs are summarized in table \ref{target}.
Among them, five objects are classified as transitional disk objects.

We reduced the HDS data in a standard manner; i.e., overscan subtraction,
bias subtraction, flat fielding, removal of scattered light, extraction
of spectra, wavelength calibration using lines of the Th-Ar lamp,
and continuum normalization.
We used IRAF packages for all procedures.
A detail description of the data reduction method
can be found in \cite{Takagi11}.
The HIRES data were reduced using the Mauna Kea Echelle Extraction (MAKEE)
package.
The UVES data were reduced using the Gasgano package.

The emission lines of the [O I] line at 6300 \AA~
are superimposed on telluric H$_{2}$O lines.
We removed the telluric lines by dividing the object spectra by 
the spectrum of S Mon,
which is a fast-rotating O-type star.

A spectrum of a TTS consists of stellar continuum,
photospheric absorption lines, forbidden emission lines, and
continuum excesses due to the boundary layer and the circumstellar disk.
To extract the forbidden emission lines from the spectra,
we removed the continuum excess and
filled the photospheric absorption lines using the spectrum of a field
dwarf with
the same spectral type.
The continuum excess veils the intrinsic photospheric absorption lines.
Flux of the stellar continuum and the strengths of the absorption lines,
thus also strengths of the forbidden emission lines, cannot be estimated
unless the amount of the continuum excess is determined precisely.
Estimation procedure of the amount of the continuum excess consists of 
several steps.
First, we estimated the radial velocities of the TTSs 
and the field dwarfs
from the wavelengths of several absorption lines around 4069 \AA,
5577 \AA, and 6300 \AA.
The spectra of the TTSs  and the field dwarfs 
were shifted, so that
their radial velocities were zero.
The widths of the absorption lines were then measured.
It is known that TTSs are often fast rotators.
The field dwarf spectra were convolved with a Gaussian profile so that the
full-width at half-maximum (FWHMs) 
of the photospheric absorption lines were comparable with those of the
TTS spectra.
We were then able to create
a veiled spectrum of a dwarf, $S'$, using the convolved
field dwarf spectrum, $S$, as follows:
\begin{equation}
S' = \frac{S+r}{1+r},
\end{equation}
where $r$ is the amount of veiling.
We calculated $r$ by comparing the equivalent widths of
absorption lines near the forbidden
lines in the TTS spectrum with those in the field dwarf spectrum.
The veiled dwarf spectrum was subtracted from the TTS spectrum, then
we added unity.
With this process, photospheric features remained for 5 TTSs 
.
For the other TTSs , 
the photospheric absorption lines were removed,
as well as continuum excess.

We measured the equivalent widths
of the forbidden lines by fitting the line profiles using Gaussian 
functions.

\begin{table}
\begin{center}
\caption{Targets}
\label{target}
\begin{tabular}{lccccl}
\hline
\hline
Object & Spectral Type & $V$-mag & $A_{\rm V}$ & 
log $\dot{\rm M}_{\rm acc}$ [M$_{\odot}$ yr$^{-1}$] & 
Telescope/Instrument \\
\hline
\multicolumn{6}{c}{Classical T Tauri Stars} \\
\hline
BP Tau   & K7    & 12.16 & 0.49 & -7.54 & Keck/HIRES \\
CI Tau   & K7    & 12.99 & 2.10 &  ---  & Subaru/HDS \\
DE Tau   & M1    & 13.04 & 0.59 & -7.59 & Subaru/HDS \\
DG Tau   & K5    & 12.43 & 1.60 & -6.30 & Subaru/HDS \\
DK Tau   & K6    & 12.35 & 0.76 & -7.42 & Keck/HIRES \\
DP Tau   & M0    & 14.22 & 1.46 & -7.88 & Keck/HIRES \\
DR Tau   & K7    & 11.61 & 1.00 & -6.50 & Subaru/HDS \\
GK Tau   & K7    & 12.54 & 0.87 & -8.19 & Keck/HIRES \\
HN Tau   & K5    & 13.85 & 0.52 & -8.89 & Keck/HIRES \\
HO Lup   & K7    & 13.00 & 1.60 & -6.74 & VLT/UVES   \\
Sz 76    & M1    & 15.18 & 1.90 &  ---  & Keck/HIRES \\
UY Aur   & K7    & 12.99 & 1.35 & -7.18 & Subaru/HDS \\
V853 Oph & M3.75 & 13.65 & 2.00 &  ---  & Keck/HIRES \\
\hline
\multicolumn{6}{c}{Transitional Disk Objects} \\
\hline
DN Tau   & K7 & 11.41 & 0.49 &  ---  & Subaru/HDS \\
GM Aur   & K7 & 12.03 & 0.14 & -8.02 & Subaru/HDS \\
LkCa 15  & K5 & 12.41 & 0.62 &  ---  & Keck/HIRES \\
TW Hya   & K7 & 11.27 & 0.00 & -8.82 & VLT/UVES   \\
V836 Tau & K7 & 13.12 & 0.59 & -9.80 & Keck/HIRES \\
\hline
\end{tabular}
\end{center}
\end{table}

\section{Results}

Figure \ref{spec} shows the forbidden line spectra 
of the classical TTSs and the transitional disk objects,
where the photospheric absorption lines were subtracted.
Some forbidden lines exhibited a low-velocity (narrow) component 
in addition to a high-velocity (broad) component.
We consider that the low-velocity component is of wind origin and the
high-velocity component is of jet origin.
Equivalent widths
of the forbidden lines 
are listed in table \ref{ewwind}.
The amount of the veiling is tabulated in table \ref{tabveil}.

\begin{table}
\begin{center}
\caption{Equivalent widths of low-velocity components of the forbidden lines}
\label{ewwind}
\begin{tabular}{cccc}
\hline
\hline
Object & 
\multicolumn{3}{c}{Equivalent width [\AA]} \\
& [S II] 4069 \AA & [O I] 5577 \AA & [O I] 6300 \AA \\
\hline
BP Tau &  0.42$\pm$0.04 & 0.20$\pm$0.03 & 0.37$^{+0.04}_{-0.03}$ \\
DG Tau &  1.33$^{+0.23}_{-0.11}$ & 0.27$\pm$0.01 & 1.70$^{+0.07}_{-0.15}$ \\
DK Tau & 0.43$^{+0.04}_{-0.05}$ & 0.07$\pm$0.01 & 0.44$\pm$0.03 \\
DP Tau & 3.80$^{+0.04}_{-0.05}$ & 0.77$^{+0.03}_{-0.04}$ & 3.90$\pm$0.10 \\
GK Tau & 0.58$^{+0.17}_{-0.14}$ & 0.08$^{+0.03}_{-0.02}$ & 0.48$^{+0.05}_{-0.06}$ \\
HN Tau & 1.30$^{+0.06}_{-0.02}$ & 0.21$^{+0.04}_{-0.08}$ & 1.24$^{+0.07}_{-0.05}$ \\
HO Lup &  0.19$\pm$0.02 & 0.29$\pm$0.02 & 0.63$^{+0.03}_{-0.02}$ \\
Sz 76   & 0.75$^{+0.24}_{-0.16}$ & 0.42$\pm$0.04 & 1.49$^{+0.09}_{-0.04}$ \\
UY Aur & 0.16$^{+0.01}_{-0.03}$ & 0.05$\pm$0.01 & 0.33$^{+0.04}_{-0.03}$ \\
V853 Oph & 2.00$^{+0.06}_{-0.03}$ & 0.44$\pm$0.05 & 2.60$^{+0.05}_{-0.06}$ \\
GM Aur & 0.07$\pm$0.02 & 0.09$^{+0.03}_{-0.02}$ & 0.38$\pm$0.02 \\
TW Hya &  0.07$\pm$0.01 & 0.07$\pm$0.01 & 0.47$^{+0.02}_{-0.01}$ \\
V836 Tau & 1.25$^{+0.08}_{-0.05}$ & 0.12$\pm$0.04 & 0.59$^{+0.05}_{-0.03}$ \\
\hline
\hline
\end{tabular}
\end{center}
\end{table}

\begin{table}
\begin{center}
\caption{Amount of continuum veiling near the forbidden lines}
\label{tabveil}
\begin{tabular}{cccc}
\hline
\hline
Object & [S II] 4069 \AA & [O I] 5577 \AA & [O I] 6300 \AA \\
\hline
\hline
BP Tau   & 2.3 & 1.0  & 0.5 \\
DG Tau   & 5.0 & 2.8  & 2.5 \\
DK Tau   & 1.4 & 0.5  & 0.3 \\
DP Tau   & 3.0 & 2.5  & 1.0 \\
GK Tau   & 1.5 & 1.0  & 0.5 \\
HN Tau   & 2.0 & 3.0  & 1.2 \\
HO Lup   & 7.5 & 3.5  & 1.6 \\
Sz 76    & 0.1 & 0.4  & 0.2 \\
UY Aur   & 9.0 & 1.8  & 1.1 \\
V853 Oph & 2.0 & 1.5  & 0.9 \\
GM Aur   & 0.9 & 0.3  & 0.3 \\
TW Hya   & 0.8 & 0.4  & 0.25 \\
V836 Tau & 0.2 & 0.25 & 0 \\
\hline
\hline
\end{tabular}
\end{center}
\end{table}

The fluxes of the forbidden lines were calculated.
We first corrected the interstellar extinction for each object.
The amounts of the extinction at the $V$-band ($A_{\rm V}$) were referred from
\cite{Geoffray01}, \cite{Kenyon95}, \cite{Hughes94},
\cite{Hamann92}, and
\cite{Kitamura96}.
Extinctions at the $B$- and $R$-bands were then calculated from that at
the $V$-band with the extinction law of \cite{Rieke85}.
Extinctions at the $B$-, $V$-, and $R$-bands were applied to the
$B$-, $V$-, and $R$-band magnitudes of the objects.
These bands correspond to the
forbidden lines of [S II] at 4069 \AA, [O I] at 5577 \AA, and
[O I] at 6300 \AA, respectively.
Assuming that continuum emission was dominant in each band,
we calculated the fluxes of the continuum level in the object spectra prior to
the subtraction of the veiled dwarf spectrum.
The fluxes of the forbidden lines were then estimated.
The derived flux ratios of the low-velocity components of
the forbidden lines after extinction
correction are listed in table \ref{ratio}.

\begin{figure}
\centering
\includegraphics[width=14.0cm, angle=0]{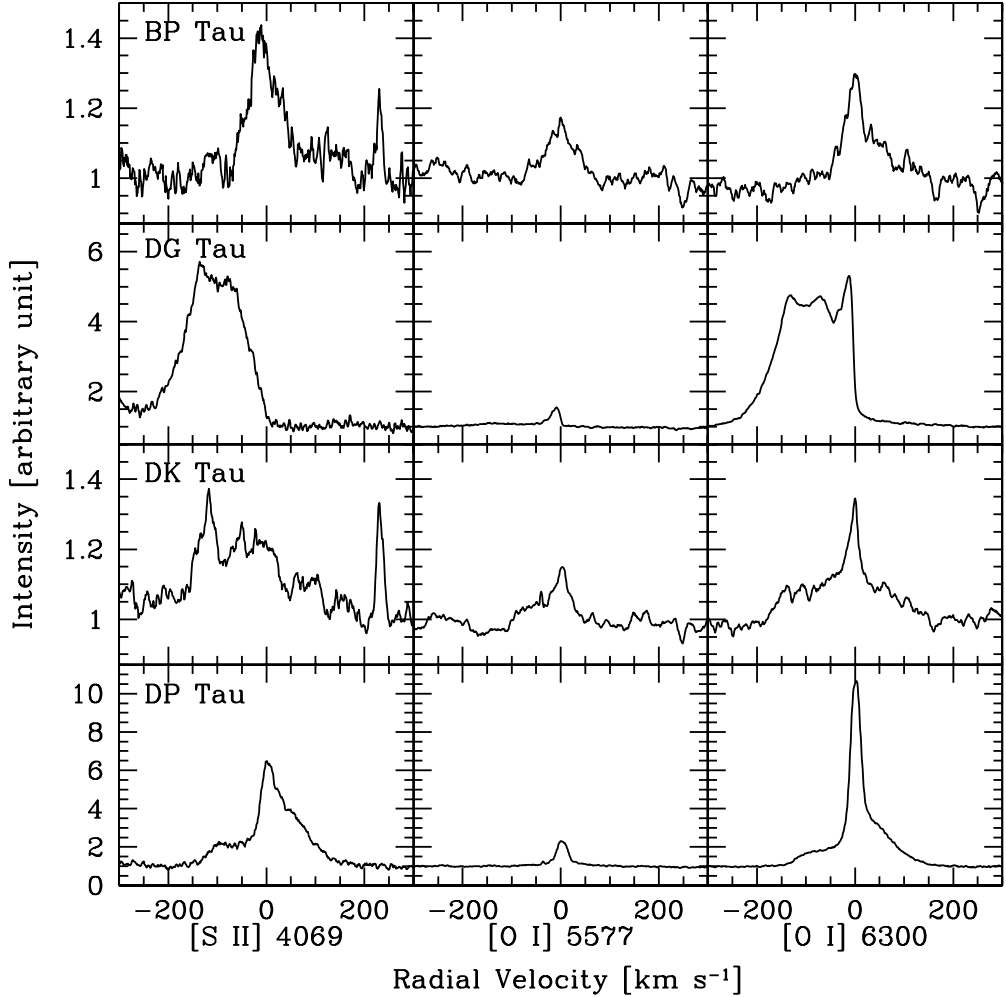}
\caption{
Forbidden emission lines of TTSs. 
Photospheric absorption lines and the continuum
excess were subtracted and the continuum levels were normalized
to unity.
}
\label{spec}
\addtocounter{figure}{-1}
\end{figure}

\begin{figure}
\centering
\includegraphics[width=14.0cm, angle=0]{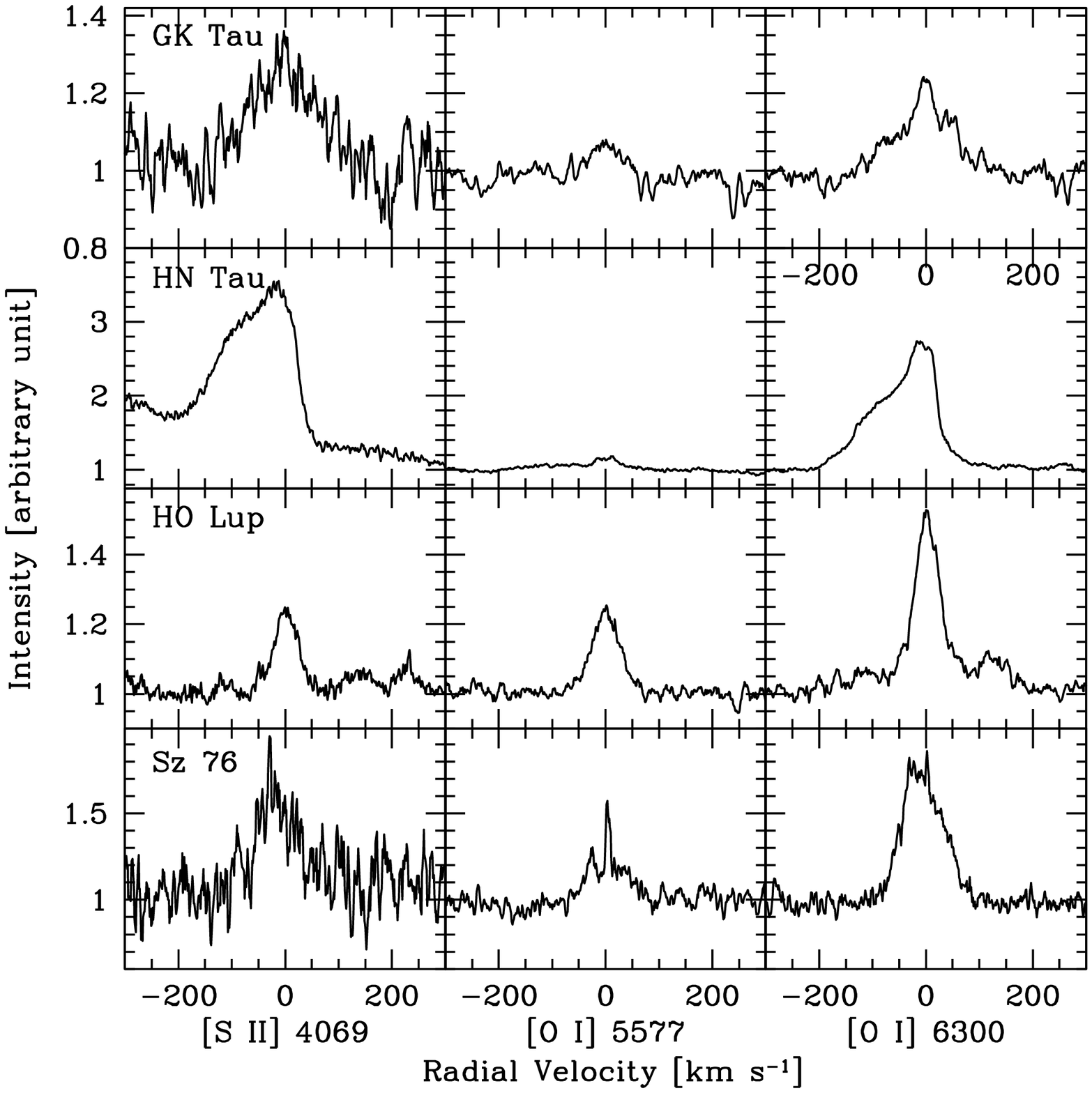}
\caption{
continued.
}
\addtocounter{figure}{-1}
\end{figure}

\begin{figure}
\centering
\includegraphics[width=14.0cm, angle=0]{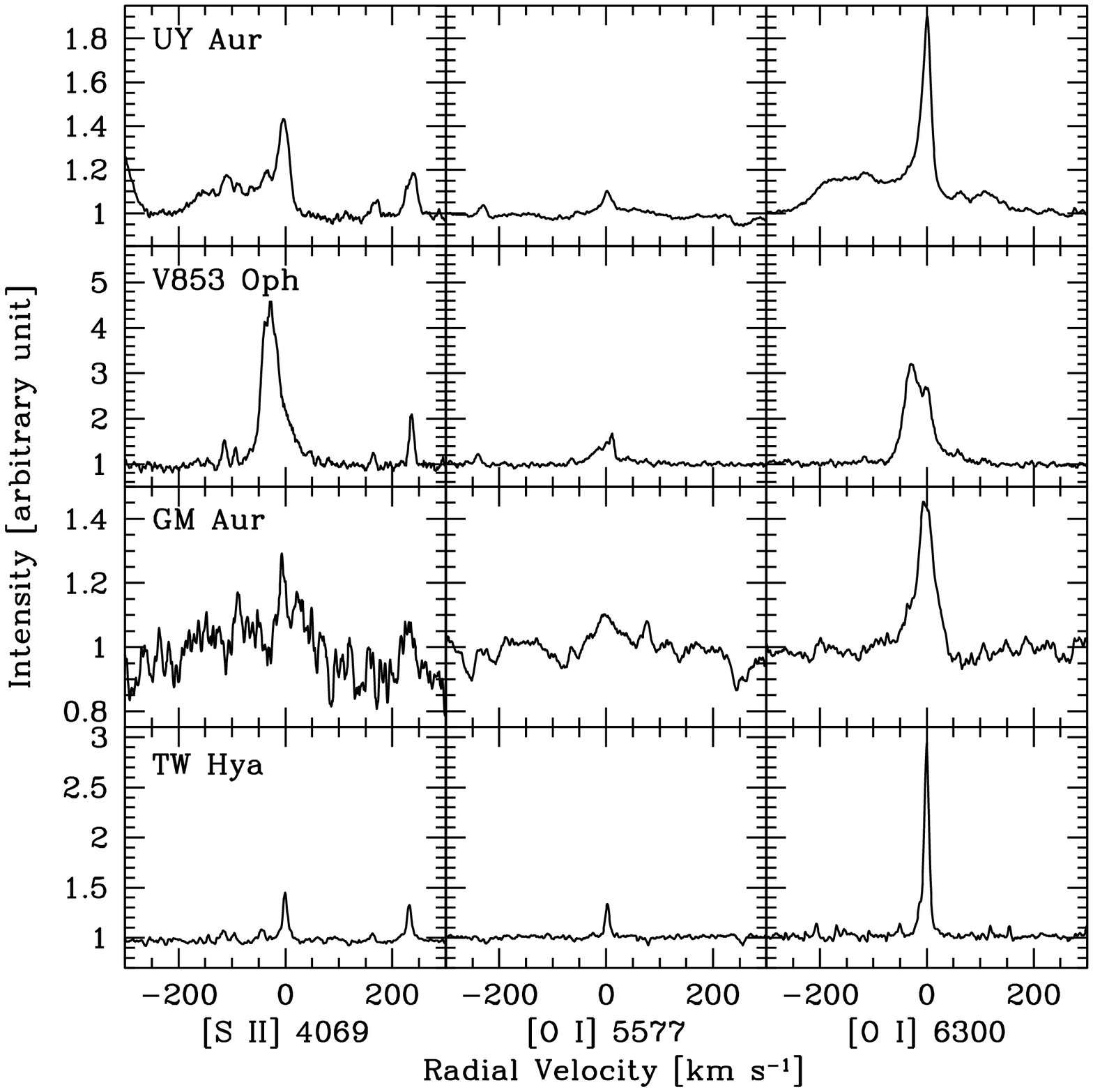}
\caption{
continued.
}
\addtocounter{figure}{-1}
\end{figure}

\begin{figure}
\centering
\includegraphics[width=14.0cm, angle=0]{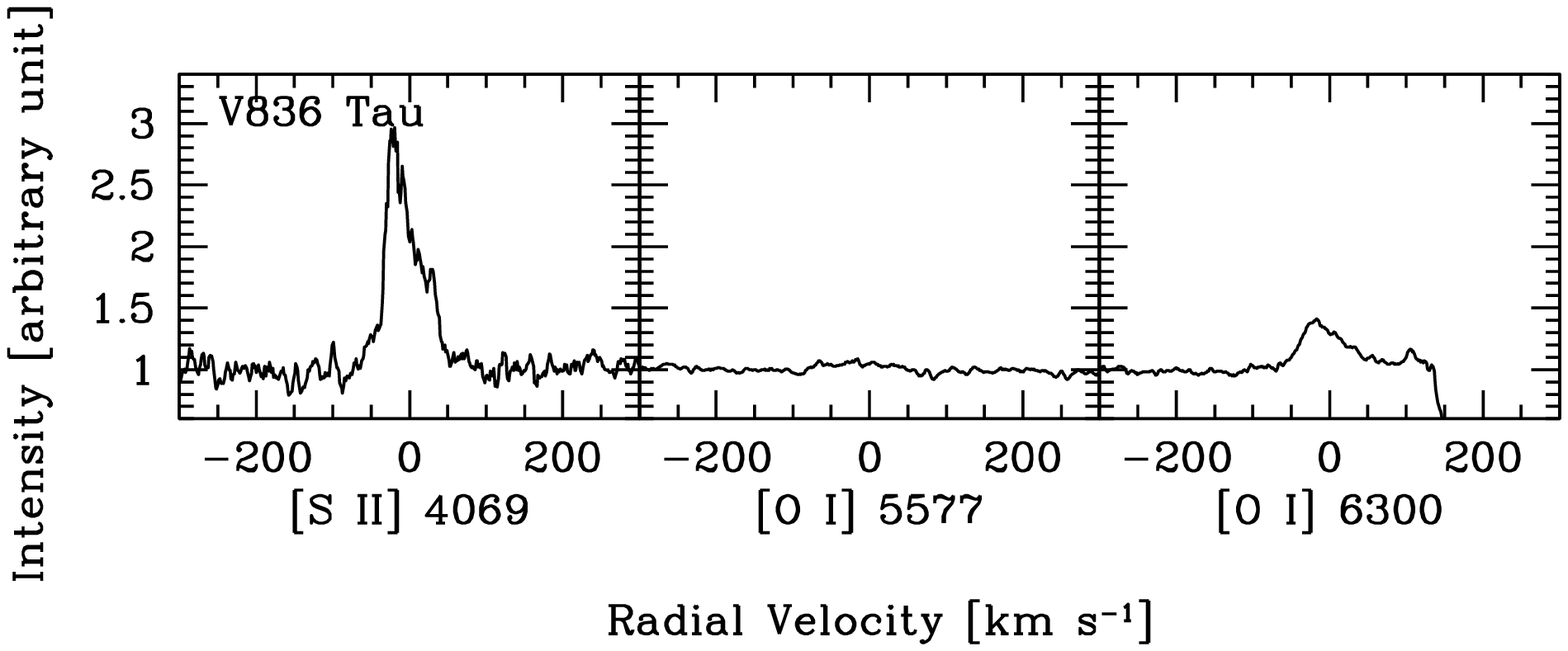}
\caption{
continued.
}
\end{figure}

\begin{table}
\begin{center}
\caption{Flux ratios of the forbidden lines}
\label{ratio}
\begin{tabular}{ccc}
\hline
\hline
Object & [S II] 4069 / [O I] 6300 & [O I] 5577 / [O I] 6300 \\
\hline
\hline
BP Tau   & 0.73$^{+0.07}_{-0.09}$ & 0.46$^{+0.07}_{-0.06}$ \\
DG Tau   & 0.88$^{+0.14}_{-0.06}$ & 0.17$\pm$0.01 \\
DK Tau   & 0.59$^{+0.05}_{-0.06}$ & 0.15$^{+0.02}_{-0.01}$ \\
DP Tau   & 0.63$^{+0.05}_{-0.03}$ & 0.17$\pm$0.01 \\
GK Tau   & 0.60$^{+0.15}_{-0.12}$ & 0.14$^{+0.04}_{-0.03}$ \\
HN Tau   & 0.72$^{+0.05}_{-0.01}$ & 0.14$^{+0.02}_{-0.04}$ \\
HO Lup   & 0.40$^{+0.08}_{-0.03}$ & 0.40$^{+0.06}_{-0.04}$ \\
Sz 76    & 0.34$^{+0.07}_{-0.05}$ & 0.20$^{+0.01}_{-0.02}$ \\
UY Aur   & 0.33$^{+0.01}_{-0.05}$ & 0.13$\pm$0.02 \\
V853 Oph & 2.88$^{+0.09}_{-0.06}$ & 0.33$^{+0.03}_{-0.02}$ \\
GM Aur   & 0.08$^{+0.03}_{-0.01}$ & 0.20$^{+0.05}_{-0.03}$ \\
TW Hya   & 0.09$^{+0.00}_{-0.01}$ & 0.09$\pm$0.01 \\
V836 Tau & 0.88$\pm$0.06          & 0.17$^{+0.03}_{-0.04}$ \\
\hline
\hline
\end{tabular}
\end{center}
\end{table}

\section{Discussion}

\begin{figure}
\centering
\includegraphics[width=14.0cm, angle=0]{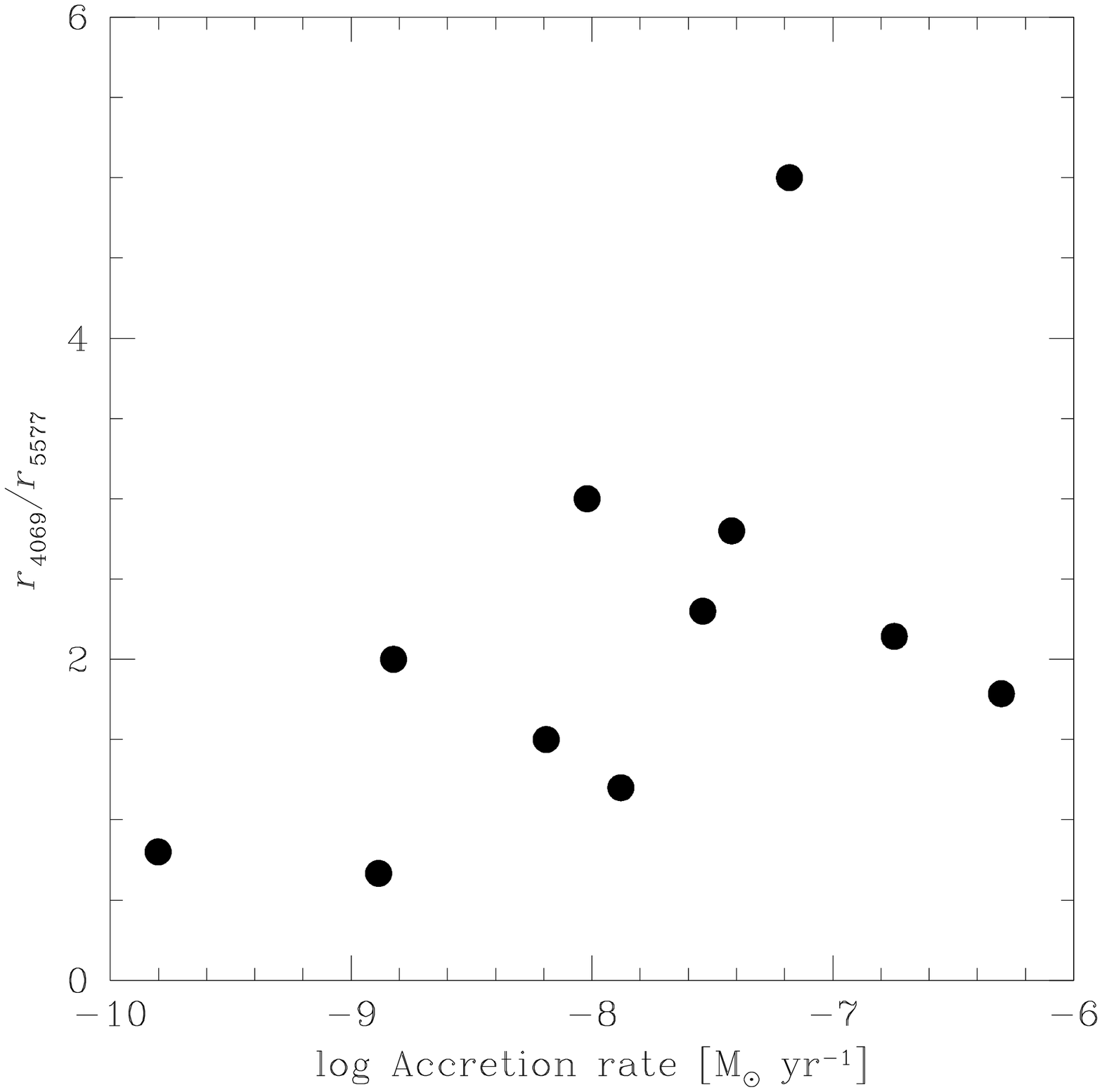}
\caption{
The ratio of the amount of the veiling at 4069 \AA~
to that at 5577 \AA~ as a function of the mass accretion rate.
The ratios increase as the temperature of the boundary layer increases.
}
\label{veiling}
\end{figure}

Veiling effect in the wavelengths shorter than the $I$-band is
attributed to mass accretion phenomena (\citealt{Bertout88}).
The continuum excess due to mass accretion exhibits
the spectral energy distribution of a blackbody with a temperature
of 8000 K -- 10000 K (\citealt{Hartigan91}).
\cite{Basri90} measured the amount of the veiling
of TTSs in the optical wavelengths.
They found that the ratio of the veiling to the photospheric
continuum increases at wavelengths shorter than 5000 \AA~ for several TTSs.
It is considered that the major source of the optical veiling
is the boundary layer between the central star and the circumstellar disk.
They also suggested that optical veiling is dependent on the
accretion rate.
We investigated correlation between the mass accretion rate of the TTSs and
the amount of the veiling.
The mass accretion rates were referred from \cite{Gullbring98}, 
\cite{Gullbring00},
\cite{Johns-Krull00}, and \cite{Herczeg08}.
In our sample, the dependence described above is not clearly seen.
Instead, there is a correlation between the color of the veiling and the 
mass accretion rates (figure \ref{veiling}). 
It is revealed that TTSs with high mass accretion rates exhibit a
high-temperature continuum excess.

We simultaneously
determined the hydrogen density and temperature of the wind
using the two flux ratios of the low-velocity components of
the three forbidden lines.
We used the software package CLOUDY (\citealt{Ferland98})
to calculate the emissivity
of the forbidden lines, given the hydrogen
density and temperature of the gas.
We assumed that the three forbidden lines emanate from the same region.
Thus, the ratios of the emissivities of the lines correspond to
the flux ratios of the lines.
We also assumed solar abundance of S and O.
The ionization parameter of the program
indicates the ratio of incident ionization photon density to
the hydrogen density of a gas.
We set it to 10$^{-15}$.
The emissivities of the [S II] 4069, [O I] 5577, and [O I] 6300
lines were calculated for the hydrogen density
between 10$^{3}$ cm$^{-3}$ and 10$^{11}$ cm$^{-3}$ with 0.2 dex intervals
and for the temperature between 6000 K and 20000 K with an interval of 
500 K.

\begin{figure}
\centering
\includegraphics[width=14.0cm, angle=0]{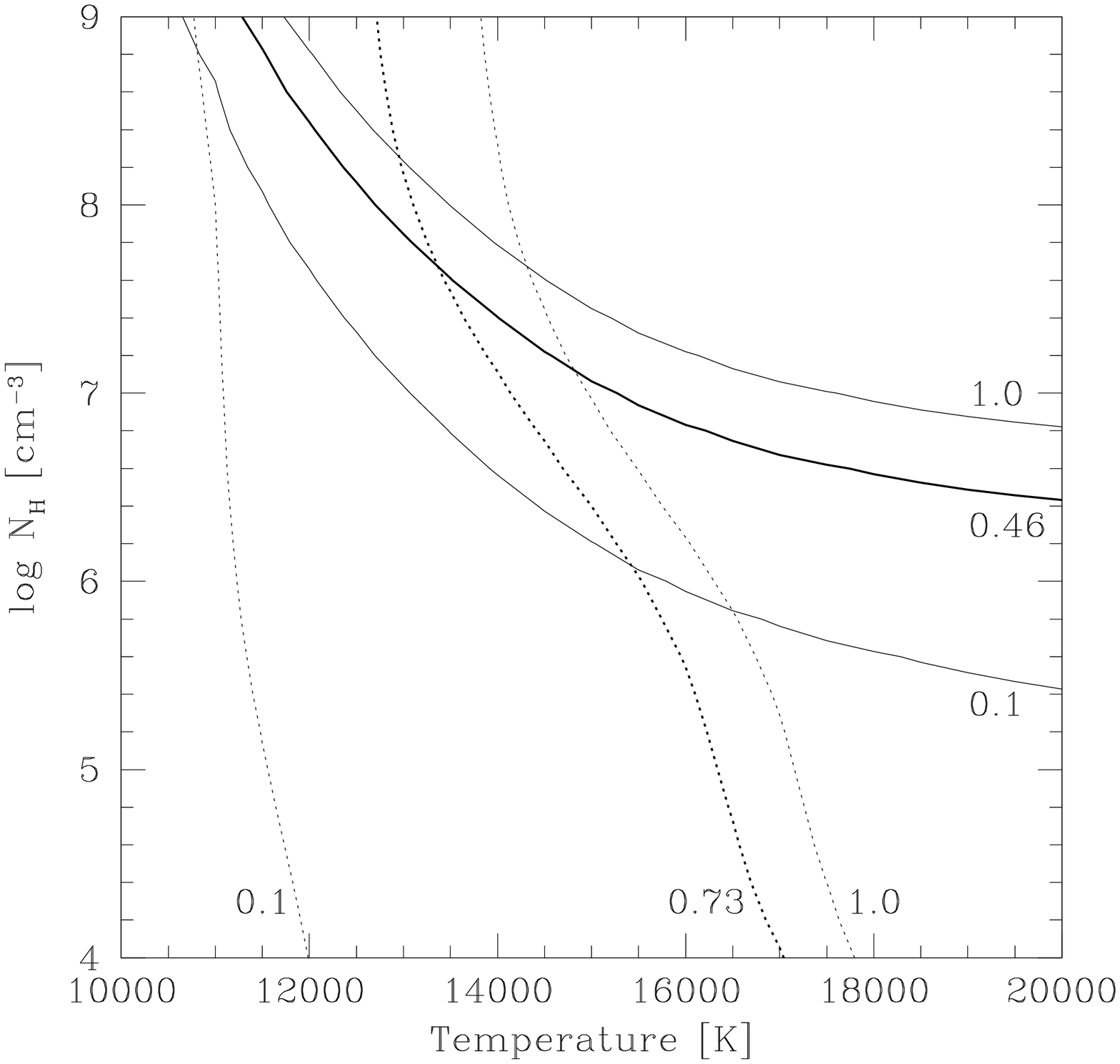}
\caption{
The flux ratios of three forbidden lines emanating from BP Tau.
The thick solid line indicates [O I] 5577 / [O I] 6300 = 0.46.
The thin lines show the ratios of 0.1 and 1.0.
The thick dotted line indicates [S II] 4069 / [O I] 6300 = 0.73,
and the thin dotted lines show the ratios of 0.1 and 1.0.
From this figure, the hydrogen density of the wind is found to be
$(6.0^{+4.0}_{-3.5})\times10^{7}$ cm$^{-3}$ and the temperature is found to be
$13000^{+1000}_{-500}$ K.
}
\label{trhobptau}
\end{figure}

As an example, the case of BP Tau is shown in figure \ref{trhobptau}.
The ratio of the observed flux of 
the [O I] 5577 line to that of the [O I] 6300 line is 0.46,
and the ratio of the observed flux of
the [S II] 4069 line to that of the [O I] 6300 line
is 0.73,
after the extinction correction.
Such ratios are reproduced when
the hydrogen density of the gas is
$(6.0^{+4.0}_{-3.5})\times10^{7}$ cm$^{-3}$ and the temperature is
$13000^{+1000}_{-500}$ K.
We find that neither the hydrogen density nor the temperature change
with the ionization parameter, unless it exceeded 10$^{-11}$.
With an ionization parameter larger than 10$^{-9}$, no combination
of parameters of the hydrogen density and temperature of the gas
could reproduce the two observed ratios of line fluxes.
In the same manner, we determined the hydrogen density and temperature of the 
wind for the 13 TTSs (table \ref{trho}).
The hydrogen densities of the winds are found to be between
2.5$\times$10$^{6}$ and  2.5$\times$10$^{9}$ cm$^{-3}$.
The wind temperatures are found to be between 10800 K
and 18000 K.
For the transitional disk objects, the density and temperature of the wind
were calculated for the three objects.
The resulting temperatures of two transitional disk
objects appear low among the targets.
However, there is no correlation between the mass accretion rate and the 
hydrogen density of the wind nor the wind 
temperature, for the whole sample.

\begin{table}
\begin{center}
\caption{Estimated hydrogen density and temperature of the winds}
\label{trho}
\begin{tabular}{ccc}
\hline
\hline
Object & log (N$_{\rm H}$) [cm$^{-3}$] & T [K] \\
\hline
\hline
BP Tau   & 7.78$^{+0.22}_{-0.38}$ & 13000$^{+1000}_{-500}$ \\
DG Tau   & 6.4$^{+0.3}_{-0.1}$    & 15000$^{+1000}_{-0}$ \\
DK Tau   & 7.78$^{+0.62}_{-1.30}$ & 12200$^{+2300}_{-700}$ \\
DP Tau   & 7.0$^{+0.3}_{-0.4}$    & 14000$\pm$500 \\
GK Tau   & 7.78$^{+1.1}_{-1.48}$  & 12500$^{+2750}_{-250}$ \\
HN Tau   & 6.48$^{+0.22}_{-0.48}$ & 15000$\pm$500 \\
HO Lup   & 8.7$^{+0.18}_{-0.3}$   & 11500$^{+500}_{-0}$ \\
Sz 76    & 8.48$^{+0.12}_{-0.48}$ & 11500$^{+500}_{-250}$ \\
UY Aur   & 8.0$^{+0.48}_{-0.15}$  & 11800$^{+200}_{-300}$ \\
V853 Oph & 6.48$^{+0.06}_{-0.08}$ & 18000$\pm$250 \\
GM Aur   & 9.4$^{+0.2}_{-0.4}$    & 10800$^{+200}_{-300}$ \\
TW Hya   & 8.78$^{+0.22}_{-0.08}$ & 11000$\pm$250 \\ 
V836 Tau & 6.48$^{+0.22}_{-0.48}$ & 15250$\pm$750 \\
\hline
\hline
\end{tabular}
\end{center}
\end{table}

\cite{Krasnopolsky03} simulated an
axisymmetric outflow which was magnetocentrifugally
driven from an inner portion of an accretion disk around a TTS.
The hydrogen density of the outflow was calculated to be 
$10^{6}$ cm$^{-3}$ at 5 AU from 
the central star and $10^{9}$ cm$^{-3}$ at the innermost region.
These densities are consistent with the densities of the wind calculated
from the observations described here.
The isodensity region of the outflow is spherical, at least for the region
with a density higher than $10^{6}$ cm$^{-3}$.

\cite{Ferro-Fontan03} investigated the thermal structure of
the wind.
The disk wind was found to have a 
temperature of $\sim$ 10000 K in the densest region (10$^{9}$ cm$^{-3}$)
and $\sim$ 15000 K in the less dense region (10$^{6}$ cm$^{-3}$).
These temperatures are consistent with the observed temperatures of the wind.
The consistency of the hydrogen density and temperature of the winds between
the observations and the simulations supports the idea that
the TTS wind emanates from the inner portion of the circumstellar disk.
Note that the hydrogen density and temperature of the winds
calculated from the observations correspond to 
the average values of the wind region; this is
because the spatial resolution of the observations is as large as
a hundred AU.

We estimated the mass and the mass loss rate of the wind.
We used the flux of the [O I] 6300 line for the mass estimates.
The fluxes of the forbidden line were translated into the luminosity
of the line, given the distance.
We assumed that the distance to the Taurus molecular cloud is 140 pc
(\citealt{Elias78}),
that to the Lupus molecular cloud is 150 pc (\citealt{Crawford00}),
that to the V853 Oph is 135 pc (\citealt{Mamajek08}),
and that to TW Hya is 54 pc (\citealt{van Leeuwen07}).
By dividing the luminosity of the line by the emissivity of the line,
one derives the volume of the emission region.
The volumes range from $3.3\times10^{37}$ cm$^{3}$
to $6.4\times10^{41}$ cm$^{3}$.
Assuming a spherical emission region, the radius was 
estimated to be between 0.22 AU and 5.7 AU.
Note that this extent is consistent with the radius of the region
from which the wind with a density of $10^{6}$ cm$^{-3}$
and higher emanates (\citealt{Krasnopolsky03}).
By multiplying this volume by the hydrogen density, 
the masses were estimated to be
between $4.7\times10^{-11}$ M$_{\odot}$ and $1.3\times10^{-9}$ M$_{\odot}$.
Assuming that the wind emanates
from the surface of the circumstellar disk
with a vertical speed of 10 km s$^{-1}$,
the mass loss rates were derived to be between 
$2.0\times10^{-10}$ M$_{\odot}$ yr$^{-1}$ and
$1.4\times10^{-9}$ M$_{\odot}$ yr$^{-1}$.
The emissivity and luminosity of the [O I] 6300 line are listed
in table \ref{mass}, as well as
the masses and the mass loss rates of the wind.

\begin{figure}
\centering
\includegraphics[width=14.0cm, angle=0]{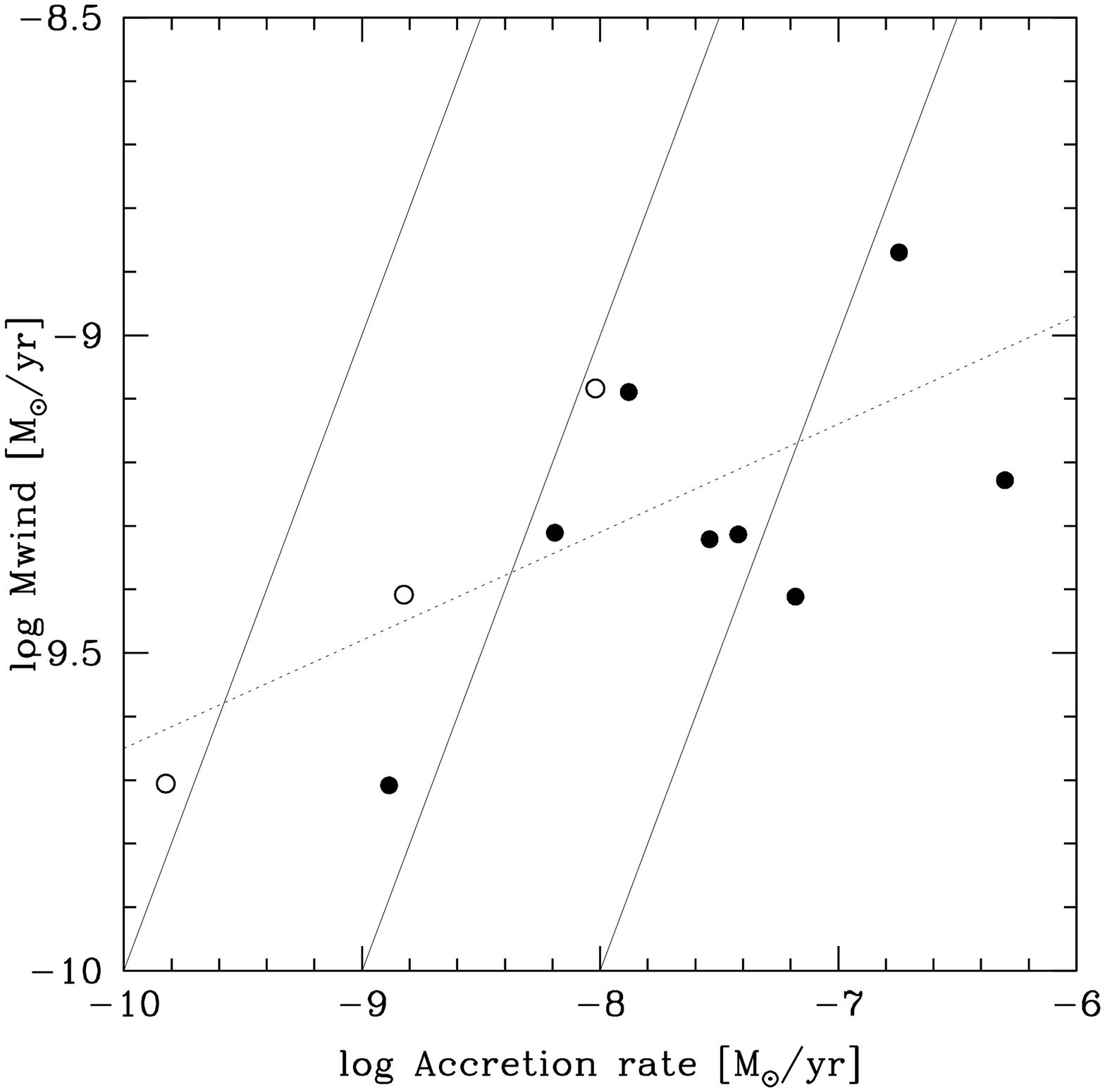}
\caption{
The mass loss rate of the wind as 
a function of the mass accretion rate from the circumstellar disk.
The filled circles represent the classical TTSs and 
the open circles represent the transitional disk objects.
The dotted line indicates the relationship between the mass loss rates
and the mass accretion rates (equation 2).
}
\label{WindAcc}
\end{figure}

Figure \ref{WindAcc} shows the relationship between the mass accretion rates
and the mass loss rates.
The ratios of the mass loss rate to the mass accretion rate are 0.001 -- 0.1
for the classical TTSs and 0.1 -- 1 for the transitional disk objects.
We claim that the wind is one of the dominant mass loss processes of the
system, at least for transitional disk objects.
The ratio of the wind mass loss rate to the accretion rate has
also been investigated from theoretical approaches and numerical
simulations.
\cite{Pelletier92} constructed
a model of a centrifugally driven hydromagnetic wind from a 
Keplerian accretion disk.
They found that the ratio of the mass loss rate to the mass accretion rate is
$\sim$ 0.1.
\cite{Sheikhnezami12} carried out numerical
simulations of a jet and an outflow from an accretion disk.
They found that 10\% -- 50 \% of the accretion material
is diverted into the jet or the wind.
These values are roughly consistent with
the ratios of the mass loss rate to the mass accretion rate
derived from the observations.

\begin{table}
\begin{center}
\caption{Estimated mass and mass loss rate of the winds}
\label{mass}
\begin{tabular}{ccccc}
\hline
\hline
Object & log $\epsilon_{\rm [O I] 6300}$ [erg s$^{-1}$ cm$^{-3}$] 
       & log L$_{\rm [O I] 6300}$ [L$_{\odot}$] 
       & log $\dot{\rm M}_{\rm loss}$ [M$_{\odot}$ yr$^{-1}$]
       & log $\dot{\rm M}_{\rm acc}$ [M$_{\odot}$ yr$^{-1}$] \\
\hline
\hline
BP Tau   & -10.7  & -4.69 & -9.3 & -7.54 \\
DG Tau   & -12    & -3.78 & -9.2 & -6.3 \\
DK Tau   & -10.55 & -4.53 & -9.3 & -7.42 \\
DP Tau   & -10.7  & -4.08 & -9.1 & -7.88 \\
GK Tau   & -10.55 & -4.53 & -9.3 & -8.19 \\
HN Tau   & -12.2  & -4.82 & -9.7 & -8.89 \\
HO Lup   & -9.6   & -4.29 & -8.9 & -6.74 \\
Sz 76    & -10.05 & -4.59 & -9.0 & --- \\
UY Aur   & -10.3  & -4.76 & -9.4 & -7.18 \\
V853 Oph & -12.4  & -4.22 & -9.2 & --- \\
GM Aur   & -8.7   & -4.76 & -9.1 & -8.02 \\
TW Hya   & -9.5   & -5.12 & -9.4 & -8.82 \\
V836 Tau & -12.2  & -4.81 & -9.7 & -9.8 \\
\hline
\hline
\end{tabular}
\end{center}
\end{table}

We also found from the observations 
that the mass loss rates gradually increase with increasing
mass accretion rates, as described by
\begin{equation}
\dot{M}_{\rm loss}{\rm [M_{\odot}/yr]} = 10^{-7.95} \times
 \dot{M}_{\rm accretion}^{0.17}{\rm [M_{\odot}/yr]}.
\end{equation}
\cite{Ferreira95} introduced the ejection index, $\xi$,
\begin{equation}
\frac{\dot{M}_{\rm loss}}{\dot{M}_{\rm accretion}} = 1 - (\frac{r_{i}}{r_{e}})^{
\xi},
\end{equation}
where $r_{i}$ is the innermost radius of the disk and $r_{e}$ is the outermost
radius.
Assuming $\frac{r_{i}}{r_{e}} = \frac{1}{30}$ as same as the case of
\cite{Sheikhnezami12}, we find $\xi =0.12$ for 
$\dot{M}_{\rm accretion} = 10^{-9}$ M$_{\odot}$ yr$^{-1}$.
\cite{Ferreira97} constrained the ejection index to $0.004 < \xi < 0.08$
for a magnetically-driven jet.
On the other hand, \cite{Sheikhnezami12} derived $0.1 < \xi < 0.5$.
The value of $\xi$ calculated from the observations is roughly consistent with
these expected values.
On the other hand, we find
$\xi = 3.5 \times 10^{-4}$ for $\dot{M}_{\rm accretion}
= 10^{-6}$ M$_{\odot}$ yr$^{-1}$.
Such a small value of $\xi$ is not predicted by the theoretical studies.
However, we noticed that some objects with a small $\xi$ have
high-velocity components of the forbidden lines.
Thus, it is considered that these objects have jets, that are the main
contributor of the mass loss process.
A calculation of the mass loss rates based on unambiguously determined 
densities and temperatures of the winds and jets for many TTSs is necessary
to constrain the ejection index, the conditions of the inner disk,
and the mechanism of the outflow phenomena.

\section{Conclusions}

We measured the fluxes of three forbidden lines of 13 TTSs
with optical high resolution spectroscopy.
If the low velocity components of the lines are in fact signatures of 
a disk wind, the following conclusions apply.

\begin{enumerate}
\item The forbidden lines of [S II] at 4069 \AA, [O I] at 5577 \AA, 
and [O I] at 6300 \AA~ were detected.
With two flux ratios of the three lines,
the hydrogen density and temperature
of the winds were determined simultaneously.
The hydrogen densities of the winds were between
$2.5\times10^{6}$ cm$^{-3}$ and $2.5\times10^{9}$ cm$^{-3}$.
The temperatures of the winds range from 10800 K to 18000 K.
\item Objects with high mass accretion rates
exhibit high mass loss rates of the winds.
The ratios of the mass loss rates of the winds
to the mass accretion rates are between 0.001 and 0.1 for the classical TTSs,
and between 0.1 and 1 for the transitional disk objects.
These ratios are roughly consistent with the ratio derived from numerical
simulations.
The wind is one of the dominant processes of the mass loss, at least,
for the transitional disk objects.
\item  The objects with a high mass accretion rate exhibit high-temperature
continuum excesses. 
\end{enumerate}

\begin{acknowledgements}
We thank the telescope staff members and operators at the Subaru Telescope.
This research has made use of the Keck Observatory Archive (KOA), 
which is operated by the W. M. Keck Observatory and the NASA Exoplanet 
Science Institute (NExScI), under contract with the National Aeronautics 
and Space Administration. 
\end{acknowledgements}

\bibliographystyle{raa}
\bibliography{windrev2}

\end{document}